\pdfoutput=1
\documentclass[12pt]{article}
\usepackage[margin=.77in]{geometry}

\usepackage{amsmath,dsfont}
\usepackage{amssymb}
\usepackage{graphicx,color}
\usepackage{subcaption}
\usepackage{slashed}

\usepackage{ifpdf}
\usepackage{hyperref}
\usepackage{cite}

\newcommand{\be}{\begin{equation}}
\newcommand{\ee}{\end{equation}}
\newcommand{\ba}{\begin{eqnarray}}
\newcommand{\ea}{\end{eqnarray}}

\newcommand{\lp}{\left(}
\newcommand{\rp}{\right)}
\newcommand{\ls}{\left[}
\newcommand{\rs}{\right]}

\newcommand{\e}{\textrm{e}}

\newcommand{\w}{\wedge}
\renewcommand{\a}{\alpha}
\renewcommand{\b}{\beta}
\renewcommand{\k}{\kappa}
\renewcommand{\O}{{\cal O}}

\newcommand{\F}{\mathcal{F}}
\newcommand{\cR}{\mathcal{R}}
\newcommand{\M}{\mathcal{M}}

\newcommand{\K}{\mathcal{K}}

\def\rmi{{\rm i}}

\newcommand{\N}{{\cal N}}
\newcommand{\T}{{\cal T}}
\newcommand{\V}{{\cal V}}

\def\ib{{\bar \imath}}
\def\jb{{\bar \jmath}}
\def\K{K{\"a}hler }

\newcommand{\rf}[1]{(\ref{#1})}

\title{{\bf \Large {  dS Supergravity from 10d }}\bigskip}

\author{Renata Kallosh,$^a$ and Timm Wrase$^b$\bigskip\bigskip\\
\small $^a$Stanford Institute for Theoretical Physics, Stanford University\\ \small Stanford, CA 94305, USA\bigskip\\
\small $^b$Institute for Theoretical Physics, TU Wien,\\ \small A-1040 Vienna, Austria\bigskip
}

\date{}

\begin{document}

\maketitle

\abstract{
\noindent
We consider flux compactification of type II string theory with local sources on SU(3)-structure manifolds. By adding pseudo-calibrated anti-$Dp$-branes wrapped on supersymmetric cycles we generalize all existing models so that the effective $d=4,$ ${\cal N}=1$ supergravity now includes a nilpotent multiplet. We present a new dictionary between string theory models and K{\"a}hler potential  $K$ and  superpotential $W$ for these dS supergravities with a nilpotent multiplet and non-linearly realized local supersymmetry. In addition to KKLT and LVS with uplifting $\overline{D3}$-branes, we have now new models with uplifting $\overline{D5}$, $\overline{D6}$, $\overline{D7}$, $\overline{D9}$-branes. The new uplifting contribution to the supergravity potential due to  Volkov-Akulov supersymmetry is universal. 
As one application of our general result, we study classical flux compactifications of type IIA supergravity and find that a previously discovered universal tachyon is now absent.

}

\newpage

 \tableofcontents{}

\section{Introduction}
The  general flux compactifications of type II supergravity on SU(3)-structure manifolds in the presence of calibrated (supersymmetric) sources and an orientifold projection gives rise to four dimensional $\N=1$ supergravity theories. The details of the compactification can therefore be codified into values of the \K potential $K$ and the superpotential $W$, and potentially D-terms, that all depend on unconstrained chiral multiplets (see for example \cite{Grana:2005jc, Blumenhagen:2006ci,Koerber:2007xk,Koerber:2010bx,Tsimpis:2016bbq} for an overview).

All such compactifications do not include the recently discovered de Sitter supergravity in four dimensions \cite{Bergshoeff:2015tra, Hasegawa:2015bza, Kallosh:2015sea, Kallosh:2015tea, Schillo:2015ssx, Kallosh:2016ndd, Ferrara:2016een, DallAgata:2016syy, Freedman:2017obq}, which has in addition to standard unconstrained chiral multiplets also a nilpotent one. To include the nilpotent multiplet into all the 4d supergravities derived from the general compactifications of the 10d supergravities that we discussed above, one can do the following:
\begin{enumerate}
\item  Keep using the most general compactification backgrounds which admit calibrated D-branes and O-planes.
\item  Keep the dictionary between the standard ${\cal N}=1$ supergravity action and the ten dimensional supergravity/string theory model.
\item  Add pseudo-calibrated anti-D-branes wrapped on supersymmetric cycles and adjust the tadpole condition accordingly.
\item  Add to the four dimensional \K potential $K$ and superpotential $W$ new terms, which we derive in this paper. These new terms include a nilpotent multiplet and therefore lead to a four dimensional dS supergravity theory.
\end{enumerate}

To explain the generality of the new results we find it convenient to use the formalism of generalized complex geometry and supersymmetry/calibration correspondence, following  \cite{Koerber:2007xk,Koerber:2010bx,Tsimpis:2016bbq} and references therein. In particular, it means that we start with SU(3)-structure manifolds that admit calibrated D-branes and O-planes, which reduce the supersymmetry via the constraint
\be
(1-\Gamma_p)\epsilon =0\,.
\label{kappa}\ee
Such a constraint follows from the $\kappa$-symmetric D-brane action when the local $\kappa$-symmetry 
\be
\delta \theta(\sigma)  = (1+\Gamma_p) \kappa(\sigma) 
\ee
is gauge-fixed as proposed in \cite{Bergshoeff:1997kr}. In this context equation \rf{kappa} is an algebraic equation defining the Killing spinor.

The condition for supersymmetry \rf{kappa}  is universal and applies to all types of branes, fundamental strings, NS5-branes, D-branes and M-branes. Thus we expect that the results of this paper apply beyond the case of anti-D-branes in SU(3)-structure compactifications. Supersymmetric (world-volume) configurations are solutions of the Born-Infeld field equations, which satisfy equation \rf{kappa} for some non-vanishing $\epsilon$. The part of the bulk supersymmetry preserved by such a configuration depends on the number of linearly independent solutions of equation \rf{kappa} in terms of  $\epsilon$.

In \cite{Gibbons:1998hm} the Killing spinor equations associated with the $\kappa$-symmetry transformations of the worldvolume brane actions were studied. It was shown that  these Dirac-Born-Infeld type systems are associated with calibrations,   and  that all the world-volume solitons associated with calibrations are supersymmetric. 

The norms of the internal two Killing spinors admitted by the compactification manifold are equal to each other, so that calibrated $Dp$-branes are admitted by the manifold. In such case, using these spinors one can construct some polyforms bilinear in spinors and use a language common to both type IIA and IIB theory. 
In particular, the existence of {\it globally defined  nowhere-vanishing spinors} allows one to construct a globally defined real two-form $J$ and a complex three-form $\Omega$ as certain bilinears of these spinors. As a result there is a very nice dictionary between the string theory models with fluxes and localized sources based on 10d supergravity, and $K$ and $W$ of the four dimensional $\N=1$ supergravity.

In \cite{Koerber:2007xk,Koerber:2010bx,Tsimpis:2016bbq} a concise way of packing this dictionary is proposed, based on pure spinors which are polyforms $\Psi_1= \Psi^{\mp}$ and $\Psi_2= \Psi^{\pm} $ for IIA/IIB. These concise formulas for $K$ and $W$ depend on these two polyforms, on the properties of the compactification manifold, on the RR potentials $C$, on the NSNS 2-form $B$ and on the dilaton. The explicit formulas for $K$ and $W$ are given for example in equations  (4.40), (4.41) in \cite{Koerber:2010bx}. They involve specific combinations of polyforms involving the Hitchin function, Mukai pairing and other objects of generalized complex geometry.

We will refer to these expressions in \cite{Koerber:2010bx} as 
\be
K_{IIA/IIB} = K(z^i, \bar z^i)\, ,   \qquad   \qquad W_{IIA/IIB} = W(z^i)\,.
\ee 
When specified to the type IIA or type IIB case, these produce the well known $K$ and $W$, which we will present in detail below in appendix \ref{app:explicit}. $K$ is a real function of the chiral multiplets $z^i, \bar z^i$ and $W$ is a holomorphic function of the chiral multiplets $z^i$. This summarizes the steps shown above as 1 and 2. Now we would like to explain our step 3. This step was actively studied in string theory for the $\overline{D3}$-brane, see for example
\cite{Lust:2008zd, Bandos:2016xyu, Aalsma:2017ulu, GarciadelMoral:2017vnz, Aalsma:2018pll}. The spontaneous breaking of supersymmetry by an $\overline{D7}$-brane in the GKP background \cite{Giddings:2001yu} was studied in \cite{Dasgupta:2016prs}.

Here we will include anti-$Dp$-branes as one of the ingredients of the string theory models in ten dimensions, with any $p$, not just $p=3$. One might worry that $Dp$-branes and anti-$Dp$-branes, when wrapped on the same cycle, are moving towards each other in the compact space and could quickly annihilate. While this is somewhat model dependent, we like to stress that our general results do not require the presence of $Dp$-branes. We can satisfy the tadpole condition using $Op$-planes, fluxes and anti-$Dp$-branes only. In such a case there are certainly many examples without perturbative instabilities, like for example setups with a single anti-$Dp$-brane, potentially even placed on top of an $Op$-plane. All anti-$Dp$-branes we include are pseudo-calibrated \cite{Lust:2008zd}, so that
\be
(1+\Gamma_p)\epsilon =0\,,
\ee
since the $\kappa$-symmetry on the world-volume of the anti-$Dp$-branes has the form
\be
\delta \theta(\sigma)  = (1-\Gamma_p) \kappa(\sigma) \,.
\ee
This means supersymmetry is  non-linearly realized on the world-volume fields and spontaneously broken. The inclusion of anti-$Dp$-branes to a string theory model in addition to $Op$-planes and maybe $Dp$-branes was viewed in the past as a compactification to $\N=0$ in $d=4$, since the anti-$Dp$-branes preserve the supersymmetry opposite to the one preserved by $Dp$-branes and $Op$-planes.

Here we will show that, in fact, one should view this step as a general way of relating string theory models, with calibrated and pseudo-calibrated branes, to four dimensional dS supergravity \cite{Bergshoeff:2015tra,Hasegawa:2015bza,Kallosh:2015sea,Kallosh:2015tea,Schillo:2015ssx,Kallosh:2016ndd,
Ferrara:2016een,DallAgata:2016syy,Freedman:2017obq}. It means that via such compactifications we obtain a supergravity action, which in addition to unconstrained multiplets has also a nilpotent one.  The nilpotent multiplet represents non-linearly realized Volkov-Akulov supersymmetry \cite{Volkov:1972jx}. The action of dS supergravity interacting with matter has a local non-linearly realized supersymmetry.
      
Our step 4 is  to give the modifications of $K$ and $W$ due to the presence of the nilpotent multiplet. 
The new action has a non-linearly realized $\N=1$ supersymmetry, which is a hallmark of dS supergravity. Our main results are the new $K$ and $W$, which depend also on a nilpotent multiplet $S$, in addition to unconstrained chiral multiplets $z^i$. They are generically of the form
\ba
K^{\rm new}(z^i, \bar z^i; S, \bar S)&=& K(z^i, \bar z^i) + K_{S\bar S}(z^i, \bar z^i) S\bar S\,,\cr
\cr
W^{\rm new} (z^i, S)&=& W (z^i ) + \mu^2 S\,.
\ea
We will show that the nilpotent field metric,  $K_{S\bar S}(z^i, \bar z^i) $ is computable: for each set of ingredients in the so-called `full-fledged string theory models' one can compute $K_{S\bar S}(z^i, \bar z^i) $ as function of the overall volume, the dilaton and the volume moduli of the supersymmetric cycles on which the anti-$Dp$-branes are wrapped. In IIB we will have four cases 
\be
\overline{D9} \,  \,  \text{on a 6-cycle}\,  ,  \qquad  \overline{D7} \,  \,  \text{ on 4-cycles}\, ,  \qquad  \overline{D5} \,  \,  \text{on 2-cycles}\, ,  \qquad \overline{D3} \,  \,  \text{ on a 0-cycle}\,.
\ee
In type IIA for SU(3) structure manifolds there are no non-trivial closed 1-forms \cite{Grana:2005jc}.  Serre duality then implies that there are no 5-forms either. Poincare duality then implies that there are no non-trivial 1- and 5-cycles that can be wrapped by a $Dp$-branes. Thus, from all potential cases 
\be
\overline{D8} \,  \,  \text{on 5-cycles}\, ,  \qquad \overline{D6}\,  \,  \text{ on 3-cycles}\, ,  \qquad \overline{D4}\,  \,  \text{on 1-cycles}\,,
\ee
only one survives
\be
\overline{D6}\,  \,  \text{on 3-cycles}\, .  
\ee  
Since the nilpotent multiplet does not have a scalar component, the new potential has an additional term but still depends on the same closed string moduli.\footnote{We are setting the open string moduli on the anti-$Dp$-branes to zero for simplicity. They can be included into the general dS supergravity using additional constrained multiplets, see for example \cite{Vercnocke:2016fbt, Kallosh:2016aep} as well as  \cite{GarciadelMoral:2017vnz}.}
The new F-term potential acquires an {\it additional nowhere vanishing positive term}, as always associated with Volkov-Akulov non-linearly realized supersymmetry
\be
V^{\rm new}(z^i, \bar z^i) = V(z^i, \bar z^i) + e^{K(z^i, \bar z^i)} |D_S W|^2\,,   
\ee
where
\be
 |D_S W|^2\equiv  D_S W  K^{S\bar S}(z^i, \bar z^i)\overline{D_{S} W}\,.
\ee
The positivity of the new term in the potential is due to the positivity of $e^{K(z^i, \bar z^i)}$ and the positivity of the nowhere vanishing 
$|D_S W|^2 $ signifying the non-linear realization of the Volkov-Akulov supersymmetry.

It is rather gratifying to see that dS supergravity might be associated with string theory models in case of all pseudo-calibrated $\overline{Dp}$-branes, which should be wrapped on supersymmetric cycles of the compactification manifolds. The well-known case of the $\overline{D3}$-brane uplift \cite{Kachru:2003aw} is not unique anymore.

\section{Type II compactifications with calibrated sources}\label{sec:standard}
In this section we review (classical) flux compactifications of type II supergravity on SU(3)-structure manifolds in the presence of calibrated (i.e. supersymmetric) sources, in particular D-branes and O-planes (see for example \cite{Grana:2005jc, Blumenhagen:2006ci} for an overview). Compactifications on SU(3)-structure manifolds give rise to four dimensional theories, which preserve linear $\N=2$ supersymmetry that is explicitly broken to linear $\N=1$ by performing in type IIA an $O6$ orientifold projection and in type IIB by performing an $O3/O7$ or $O5/O9$ orientifold projection.

The theories  that lead upon compactification on an SU(3)-structure manifold to a \emph{standard} 4d $\N=1$ theory have an action that consists of three parts, the closed string type II action, the $Op$-plane action and the $Dp$-brane action\footnote{Here $S_{Op/Dp}$ denote the action for a single plane/brane. The O-planes or D-branes can wrap different cycles but we omit a corresponding index. In the case of $O3/O7$ and $O5/O9$ there are two different $p$'s and our argument goes through in the same way.}
\be
S = S_{II} + N_{Op} S_{Op} + N_{Dp} S_{Dp}\,,
\ee
We will now split each of the above three terms into two parts, the second of which is relevant for the tadpole cancellation condition
\be
S = \tilde{S}_{II} + \int C_{p+1} \w (d F_{8-p} - H\w F_{6-p}) +N_{Op}( S^{\rm DBI}_{Op} + S^{\rm CS}_{Op})+ N_{Dp}(S^{\rm DBI}_{Dp}+ S^{\rm CS}_{Dp})\,.
\ee
Now we have that $S^{\rm CS}_{Op} = -2^{p-5} \int C_{p+1}$ and $S^{\rm CS}_{Dp} = \int C_{p+1}+\ldots$, where \ldots includes other bosonic and fermionic terms (for the ease of presentation we temporarily set the $Dp$-branes tension to one). Varying the action with respect to $C_{p+1}$ leads to the following (integrated) tadpole cancellation condition
\be\label{eq:tadpole}
\int d F_{8-p} - H\w F_{6-p}=  -2^{p-5} N_{Op} + N_{Dp}\,.
\ee
Once we satisfy this tadpole cancellation condition the remaining part of the action that gives rise to a standard 4d $\N=1$ supergravity action is
\be\label{eq:stSUGRA}
S_{\rm standard-SUGRA} = \tilde{S}_{II} +N_{Op} S^{\rm DBI}_{Op} +N_{Dp} S^{\rm DBI}_{Dp}\,.
\ee
The resulting 4d $\N=1$ supergravity action has the form 
\ba
S_{\rm standard-SUGRA}&=& \int d^4x\sqrt{-g} \lp \frac12 R - K_{i\ib}\partial_\mu z^i \partial^\mu \bar z^{\ib} - V \right.\cr
&&\qquad \qquad \quad \left.-\frac12 (\text{Re}f_{\alpha\beta}) F_{\mu\nu}^\alpha F^{\mu\nu,\beta}-\frac12 (\text{Im}f_{\alpha\beta}) F_{\mu\nu}^\alpha \tilde{F}^{\mu\nu,\beta}\rp\,.
\ea
Here the scalar potential is a combination of F-term and D-term parts and is given by
\be
V=e^K\lp K^{i\jb}D_i W \overline{D_j W}-3|W|^2 \rp +\frac12 (\text{Re} f)^{-1 \alpha \beta} D_\alpha D_\beta\,.
\label{pot}\ee
This action is determined by the real K\"ahler potential $K$, the holomorphic superpotential $W$ and the holomorphic gauge kinetic function $f_{\alpha\beta}$. These depend on the complex scalar fields $z^i$ that arise from dimensionally reducing the metric as well as the other ten dimensional string fields. As mentioned above, for an SU(3)-structure manifold we can use the Killing spinors to construct a K\"ahler (1,1)-form $J$ and a holomorphic (3,0) form $\Omega$ (see \cite{Grana:2005jc} for details). These contain the K\"ahler and complex structure moduli. Additionally, we get in the NSNS sector scalar fields from the Kalb-Ramond field $B$ and the dilaton $e^\phi$. The parameters that enter the scalar potential from the NSNS sector are the H-flux as well as so called metric fluxes that encode the curvature of the SU(3)-structure manifold. The scalars and parameters that arise in the RR-sector depend on whether we are studying type IIA or type IIB and on the particular orientifold projection. We will discuss them in detail in appendix \ref{app:explicit}.

Generically, the effective scalar potential derived from 10d for compactifications with a warped metric is given for example in eq. (4.4) in \cite{Lust:2008zd}, see the notation there. Namely the density of the 4d potential consists of 2 parts, the one from the classical 10d supergravity action with fluxes
\be
\tilde \V_{\rm eff} = \int_{\M_6} d{\rm Vol}_6 e^{4A} \Big \{ e^{-2\phi} [ -\cR_6+{1\over 2}H^2 - 4 (d\phi)^2 + 8 \nabla ^2 A+ 20 (dA)^2] -{1\over 2} \tilde F^2 \Big \}
\ee
and the part from the local sources
\be
\V_{\rm eff}^{\rm loc}= \sum_i T_i \Big ( \int _{\Sigma_i} e^{4A-\phi} \sqrt{ \det (g |_{\Sigma_i}+ \F_i} - \int _{\Sigma_i} C^{\rm el} |_{\Sigma_i} \wedge e^{\F_i} \big)
\label{luca}\ee
Here the localized sources are D-branes and O-planes, where for the O-planes we have to set ${\cal F}_i=0$. As above for $\sqrt{\alpha'}=1/2\pi$ one has $T_{Dp}=1$, $T_{Op}=- 2^{p-5}$. Many examples of this setup and relations to the concept of calibrated D-branes can be found in \cite{Lust:2008zd}. We present the relevant cases in the next sections.

This concludes a short review of flux compactifications of type II supergravity on SU(3)-structure manifolds in the presence of calibrated (i.e. supersymmetric) sources (see for example \cite{Grana:2005jc, Blumenhagen:2006ci} for more details).

\section{Adding pseudo-calibrated  anti-$Dp$-branes}

In most string theory compactifications with phenomenological applications the goal was to find the ingredients of standard 4d $\N=1$ supergravity, i.e. to find $K$ and $W$ for unconstrained chiral multiplets and to identify the potential \rf{pot} associated with `full-fledged string theory models'. In \cite{Lust:2008zd} an important step was made to accommodate the KKLT  construction in this setting. At that time adding an anti-$D3$-brane, even pseudo-calibrated, meant that supersymmetry of the kind available in standard supergravity becomes broken down to $\N=0$. The additional term in the potential, the so called uplifting term in KKLT, $V_{\rm up} ={8 D\over (\rm {Im} \rho)^3}$, was not part of the potential in \rf{pot} and only the bosonic term $V_{up}$ was presented.

Since then the manifestly supersymmetric version of the KKLT uplifting was proposed in the form in which the anti-$D3$-brane is represented by a nilpotent multiplet $S$ with $S^2=0$, corresponding to VA non-linearly realized supersymmetry \cite{Kallosh:2014wsa}. In this case the new $K$ and $W$ are (in the  unwarped case) given by
\be
\begin{aligned}\label{KKLT}
&K= -3\log\left(T+\bar T\right)+S \bar S \, , \\ 
 &W= W_0 + A \exp(-a T) + \mu^2\ S\,, \qquad \Rightarrow \qquad V_{up}= \left. e^K |D_S W|^2 \right|_{S=\bar S=0}= {\mu^4\over (T+\bar T )^3}\,.
\end{aligned}
\ee
The reason why in the KKLT case the presence of $D3$-branes and $O3$-planes which were constrained by a tadpole condition, was not leading to an uplift term, is due to the fact that these were calibrated: they preserved the same symmetry as the background, $(1-\Gamma_p)\epsilon =0$. Meanwhile, the anti-$D3$-branes are pseudo-calibrated, they preserve the symmetry opposite to that of the background and the $D3$-branes/$O3$-planes, $(1+\Gamma_p)\epsilon =0$.

The concept of calibrated $Dp$-branes and pseudo-calibrated pseudo-calibrated anti-$Dp$-branes is totally general. From this perspective, in dS supergravity  constructions there is no need to restrict ourselves to anti-$D3$-branes as an exclusive source of Volkov-Akulov non-linearly realized supersymmetry. Any D-brane has a non-linearly realized supersymmetry and therefore one has to look at the general case of including pseudo-calibrated anti-$Dp$-branes, wrapped on supersymmetric cycles, as new local sources, and check the tadpole condition, as suggested in point 3 in the Introduction.

From all possible $Dp$-branes with $p\geq 3$ we can get uplift terms, i.e. positive new terms in the 4d scalar potential, if there are supersymmetric $(p-3)$-cycles on our compactification manifold. In type IIB there are 6-, 4-, 2-, 0-cycles, therefore we will have an uplift term due to anti-$D9$-, anti-$D7$-, anti-$D5$-, anti-$D3$-branes. In type IIA on SU(3)-structure manifolds there are only 3-cycles and therefore only anti-$D6$-branes can give rise to a new positive uplift term in the scalar potential.

Let us now repeat the general derivation of the four dimensional action at the beginning of section \ref{sec:standard} but now we also include anti-$Dp$-branes. The action is 
\be
S = S_{II} + N_{Op} S_{Op} + N_{Dp} S_{Dp} + N_{\overline{Dp}} S_{\overline{Dp}}\,,
\ee
We again split each of the above terms into two parts and use that $S_{\overline{Dp}} = S^{\rm DBI}_{Dp}- S^{\rm CS}_{Dp}$
\be
S = \tilde{S}_{II} + \int C_{p+1} \w (d F_{8-p} - H\w F_{6-p}) +N_{Op}( S^{\rm DBI}_{Op} + S^{\rm CS}_{Op})+ N_{Dp}(S^{\rm DBI}_{Dp}+ S^{\rm CS}_{Dp}) + N_{\overline{Dp }}(S^{\rm DBI}_{Dp}- S^{\rm CS}_{Dp})\,.
\ee
Varying the action with respect to $C_{p+1}$ leads now to the following (integrated) tadpole cancellation condition
\be\label{eq:newtadpole}
\int d F_{8-p} - H\w F_{6-p}=  -2^{p-5} N_{Op} + N_{Dp} - N_{\overline{Dp}}\,.
\ee
Once we satisfy this tadpole cancellation condition the remaining part of the action that now will give rise to a new 4d $\N=1$ dS supergravity action is
\be
S_{\rm dS-SUGRA} = \tilde{S}_{II} +N_{Op} S^{\rm DBI}_{Op} +N_{Dp} S^{\rm DBI}_{Dp} +N_{\overline{Dp}} S^{\rm DBI}_{Dp}\,.
\ee
The above action is actually related to the standard supergravity action in equation \eqref{eq:stSUGRA} in a very simple way. Let us assume for example that we satisfy the new tadpole condition in equation \eqref{eq:newtadpole} by not changing the fluxes on the left-hand-side nor $N_{Op}$, but simply by adding an additional $N_{\overline{Dp}}$ $Dp$-branes so that $N_{Dp} \rightarrow N_{Dp}+N_{\overline{Dp}}$. Then we find that the new action has the form 
\be
S_{\rm dS-SUGRA} = S_{\rm standard-SUGRA} + 2 N_{\overline{Dp}} S^{\rm DBI}_{Dp}\,.
\ee
So the new action is actually related to the old one by adding twice the DBI action for the anti-$Dp$-brane. This result holds in full generality also in the absence of any $Dp$-branes. In this case one has to adjust the fluxes because of the tadpole condition in equation \eqref{eq:newtadpole}. This adjustment of the fluxes then modifies $\tilde{S}_{II}$ exactly in the right way to give the new term in the dS supergravity action. 

Therefore, for all anti-$Dp$-branes we find that they lead to a new contribution to the scalar potential in four dimensions that is in string frame of the form
\be\label{eq:newterm}
V_{\overline{Dp}} = 2 N_{\overline{Dp},\alpha} T_{Dp} \int_{\Sigma_\alpha} d^{p-3}\xi e^{-\phi} \sqrt{\text{det}( G+B-2\pi \a' F)}\,,
\ee
where $\a$ labels the different $(p-3)$-cycles $\Sigma_\a$ that are wrapped by the anti-$Dp$-branes and $T_{Dp}$ denotes their tension. In the next two sections we will work out exactly how this new term can be included in the K\"ahler and superpotential via a nilpotent chiral superfield. For simplicity we do not include the worldvolume scalar fields on the anti-$Dp$-branes, like the gauge field or the position moduli in our discussion. It should be possible to include them using other constrained multiplets as in \cite{Vercnocke:2016fbt, Kallosh:2016aep}. Note however that these moduli could be absent in some cases, if we for example place a single anti-$Dp$-brane on top of an $Op$-plane.

In all cases we will find that
\be
V^{\rm new}(z^i, \bar z^i) = V(z^i, \bar z^i) + e^{K(z^i, \bar z^i)} D_S W  K^{S\bar S}(z^i, \bar z^i)\overline{D_{S} W}
\ee
and the dictionary between string theory models with anti-$Dp$-branes and dS supergravity with a nilpotent multiplet will be established.

\subsection{Pseudo-calibrated anti-$Dp$-branes in type IIB}
The calibration condition for $p=3, 5, 7, 9,$ is given in \cite{Blumenhagen:2006ci} in the paragraph between (2.185) and (2.186). It allows us to rewrite the new positive term in the scalar potential, given above in \eqref{eq:newterm}, as
\be
V_{\overline{Dp}} =2 N_{\overline{Dp},\a} T_{Dp} \int_{\Sigma_\a} d^{p-3}\xi e^{-\phi} \text{Re}\lp e^{J+i B}\rp\,.
\ee
Explicitly this means (see appendix \ref{app:explicit} for our notation)
\ba
V_{\overline{D3}} &=& 2 N_{\overline{D3}} T_{D3} \int d^0\xi e^{-\phi}= 2 N_{\overline{D3}} T_{D3} \text{Im}(\tau)\,,\cr
V_{\overline{D5}} &=& 2 N_{\overline{D5},\a} T_{D5} \int_{\Sigma_\a} d^2\xi e^{-\phi} J = 2 N_{\overline{D5},\a} T_{D5} \text{Im}(t^\a)\,,\cr
V_{\overline{D7}} &=& 2 N_{\overline{D7},\a} T_{D7} \int_{\Sigma_\a} d^{4}\xi \frac12 e^{-\phi}\lp J\w J- B\w B\rp =  -2 N_{\overline{D7},\a} T_{D7} \text{Im}(T^\a)\,,\cr
V_{\overline{D9}} &=& 2 N_{\overline{D9}} T_{D9} \int_{X} d^{6}\xi  e^{-\phi} \lp \frac16 J\w J\w J-\frac12 J\w B \w B\rp= -2 N_{\overline{D9}} T_{D9} \text{Im}(\T)\,.
\ea
So we see that there is a nice unifying description.

Now we go to the 4d Einstein frame.\footnote{The 10d action in string frame contains the term $S \supset \int d^{10}x \sqrt{-g^s_{10}} e^{-2\phi} R_4 = \int d^{4}x \sqrt{-g^s_4} R_4 \V_6e^{-2\phi} = \int d^{4}x \sqrt{-g^E_4} R_4$. Here we have only redefined the 4d metric $g^s_{4} \rightarrow e^{2 \phi} g_4^E/\V_6$.} Above we have already identified the correct moduli in Einstein frame so that this rescaling changes all the above expressions only due to the $\int d^4x \sqrt{-g_4^s}=\int d^4x \sqrt{-g_4^E}\frac{e^{4\phi}}{\V_6^2}=\int d^4x \sqrt{-g_4^E}e^{4\phi_4}$ factor in the DBI action. Here we defined in the last equation the four dimensional dilaton $\phi_4=\phi-\frac12\ln(\V_6)$. 

For the anti-$D3$-brane this gives the usual (unwarped) expression, if we use that for a single K\"ahler modulus in 4d Einstein frame we have $2^3 \V_6^2 =\rmi e^{3\phi}(T- \bar T)^3$,
\be
V_{\overline{D3}} =  2 N_{\overline{D3}} T_{D3} \text{Im}(\tau)\frac{e^{4\phi}}{\V_6^2} = 2 N_{\overline{D3}} T_{D3}\frac{e^{3\phi}}{\V_6^2} =2 N_{\overline{D3}} T_{D3} \frac{2^3}{\rmi(T- \bar T)^3}\,.
\ee

For all cases we have simply
\ba
V_{\overline{D3}} &=&  -\rmi T_{D3} N_{\overline{D3}} (\tau-\bar\tau)e^{4\phi_4} \,,\cr
V_{\overline{D5}} &=& -\rmi  T_{D5}N_{\overline{D5},\a} (t^\a-\bar t^\a )e^{4\phi_4} \,,\cr
V_{\overline{D7}} &=& \rmi  T_{D7} N_{\overline{D7},\a} (T^\a-\bar T^\a)e^{4\phi_4} \,,\cr
V_{\overline{D9}} &=& \rmi T_{D9} N_{\overline{D9}}  (\T-\bar\T)e^{4\phi_4} \,.
\ea
Let us introduce the shorthand notation for all cases above
\be
V_{\overline{Dp}} =2 T_{Dp} e^{4\phi_4} \text{Im}{\Phi}\,,
\ee
where Im$\Phi=\{N_{\overline{D3}} \text{Im}\tau, N_{\overline{D5},\a} \text{Im}t^\a, -N_{\overline{D7},\a} \text{Im}T^\a, -N_{\overline{D9}} \text{Im}\T\}$ is a positive, real linear combination of the respective complex moduli in the particular setups.

We can then obtain the above expression $V_{\overline{Dp}} $ from
\ba
K&=& K_{\rm before}  +\rmi  e^{K_{\rm before}} \frac{e^{-4\phi_4}}{(\Phi-\bar\Phi)} S\bar S\,,\cr
W&=&W_{\rm before} +  \mu^2 S\,,
\ea
where $\mu^4=T_{Dp}$. For the particular case of an anti-$D3$-brane this agrees with the previously derived equation (3.40) in \cite{GarciadelMoral:2017vnz}.

\subsection{Pseudo-calibrated anti-$Dp$-branes in type IIA} \label{ssec:antiD6}
Spacetime filling $Dp$-branes in type IIA wrap an odd dimensional internal cycle, this leaves us only with the case of anti-$D6$-branes, since there are no non-trivial 1- and 5-cycles.

The calibration condition for $D6$-branes is given in \cite{Blumenhagen:2006ci} in equation (2.184). It allows us to rewrite the new term in the scalar potential, given above in \eqref{eq:newterm}, as \footnote{We use slightly different conventions compared to \cite{Blumenhagen:2006ci}. We take $\rmi \int \Omega \w \bar\Omega =1$ (see eqn. (2.12) of \cite{DeWolfe:2005uu}) instead of \cite{Blumenhagen:2006ci} where $\rmi \int \Omega \w \bar\Omega = \V_6$. Hence we have an extra factor of $\sqrt{\V_6}$. As discussed above, we get an extra factor $\frac{e^{4\phi}}{\V_6^2}$ from going to 4d Einstein frame.}
\be\label{eq:VantiD6}
V_{\overline{D6}} = \frac{2 N_{\overline{D6},K} T_{D6}}{\V_6^2} \int_{\Sigma_K} e^{3\phi} \sqrt{\V_6}\text{Re}\Omega = 2 N_{\overline{D6},K} T_{D6} e^{4\phi_4} \text{Im}(Z^K)\,.
\ee
We again can write this new term by including a nilpotent chiral multiplet $S$ coupled to the other fields. In particular, one finds that
\ba
K&=& K_{\rm before} +\rmi e^{K_{\rm before}}\frac{e^{-4\phi_4}}{N_{\overline{D6},K}(Z^K-\bar{Z}^K)}S \bar S\,,\cr
W&=& W_{\rm before} + \mu^2 S\,.
\ea
where $\mu^4=T_{D6}$.

\section{dS vacua in type IIA dS supergravity}
We are now focusing on the particular case of massive type IIA flux compactification to which we can add anti-$D6$-branes as explained in subsection \ref{ssec:antiD6}. This case is particularly  simple since all moduli can be stabilized (see \cite{Danielsson:2011au} for a review of this particular class of compactifications). However, it has never been possible to find (meta-) stable dS vacua in this context. All example of dS critical points have always had at least one tachyonic direction with large slow-role parameter $|\eta| \gtrsim \O(1)$ \cite{Danielsson:2012et}. This had lead people to investigate whether there are no-go theorem's in this case that forbid stable dS vacua \cite{Shiu:2011zt, Danielsson:2012et, Junghans:2016uvg, Junghans:2016abx}. Two for us important insights have emerged from these studies: 1) The obstinate tachyonic direction involves the 3-cycle moduli \cite{Danielsson:2012et, Junghans:2016uvg, Junghans:2016abx}. 2) In the limit of very small positive value of the potential, the tachyonic direction seems be connected to the sGoldstino \cite{Covi:2008ea, Danielsson:2012et, Junghans:2016uvg, Junghans:2016abx}. 

Our new term that appears in the action does involve the 3-cycles since we can wrap them with anti-$D6$-branes, so the new term should have an effect on the tachyonic direction. Furthermore, since the anti-$D6$-branes break supersymmetry they will modify the sGoldstino direction. For dominant SUSY breaking from anti-$D6$-branes, the Goldstino will be the worldvolume fermion on the anti-$D6$-brane, which is encoded in the nilpotent field $S^2=0$. This Goldstino has no scalar partner and therefore there is no sGoldstino that is at the risk of being tachyonic.\footnote{The supersymmetry on the anti-$D6$-branes is non-linearly realized. The fermions on the worldvolume do not simply get mapped into a boson under these transformations (see for example \cite{Kallosh:2016aep} for more details).}  The explicit no-go theorem \cite{Junghans:2016abx} that predicts a tachyonic field with $\eta \leq -\frac43$ in standard type IIA compactifications is circumvented in the more general dS supergravity, due to the presence of anti-$D6$-branes.

\begin{figure}
    \centering
        \includegraphics[width=.6\textwidth]{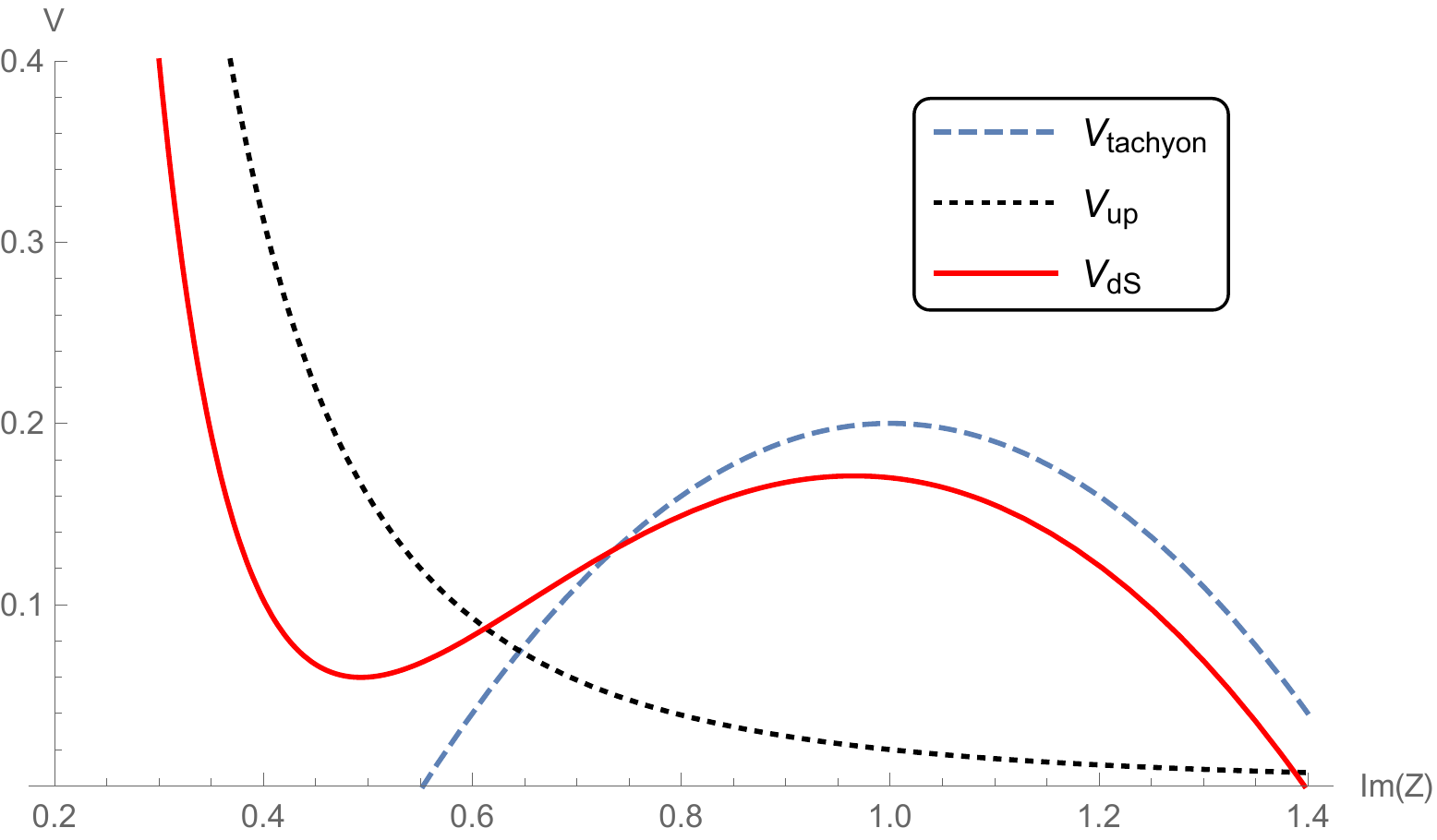}
\caption{The extra term from an anti-$D6$-brane can lead to a stable minimum near a dS saddle point. The dashed blue potential $V_{\rm tachyon} $ with the maximum is the standard potential when only calibrated D6-branes and O6-planes are present. The uplift due to the anti-D6-brane, $V_ {\rm up} \propto 1/\text{Im}(Z)^3$, is shown by a black dotted line. Finally, the sum of the original tachyonic potential and the uplift potential is shown as solid red line and exhibits a dS minimum.}
  \label{fig:uplift}
\end{figure}

Given the above it might not be guaranteed that the tachyonic direction can be absent in these models, however, we will provide a simple intuitive reason for why this is actually the case. Let us restrict to the case of a model with a single 3-cycle modulus Im$(Z)$ (or more generally this Im$(Z)$ could be the linear combination of 3-cycle moduli that is tachyonic). Then near the dS saddle point at $\text{Im}(Z)=\text{Im}(Z_0)$ the potential without the anti-$D6$-branes has the form $V_ {\rm tachyon} \propto V_0 - (\text{Im}(Z)-\text{Im}(Z_0))^2$, for some $V_0>0$. The positive new term from the anti-$D6$-branes in equation \eqref{eq:VantiD6} above has an implicit Im$(Z)$ dependence from $e^{4\phi_4} \propto 1/\text{Im}(Z)^4$, so that it scales like $V_ {\rm up} \propto 1/\text{Im}(Z)^3$. The combination of these terms then generically has a dS minimum for an appropriately chosen number of anti-$D6$-branes, as is shown in figure \ref{fig:uplift}.

We study this for the simplest known example and find indeed for appropriately tuned parameters that the obstinate tachyon is absent. In the truncation to left-invariant fields, there is no other tachyon, so it is possible that this is the first stable dS vacuum in this context. In order to know for sure, one has to check that there are no other light fields that have a negative mass, see section 3.1 of \cite{Roupec:2018fsp} for a discussion of this point.

One might worry that anti-D6-branes could annihilate quickly against the background fluxes \cite{Danielsson:2016cit}. However, the analysis in \cite{Danielsson:2016cit} is only valid for a large number of anti-D6-branes, while for a small number a different result seems likely \cite{Michel:2014lva}. So it is plausible that uplifting leads to long lived dS vacua, if one uses a single anti-D6-brane or an anti-D6-brane on top of an O6-plane.

In this simplest model the unstable dS vacua that were previously found in \cite{Caviezel:2008tf, Danielsson:2010bc} can be shown to all lie at small volume and large string coupling \cite{Danielsson:2011au} so that one expects large $\a'$ and string loop corrections. The anti-$D6$-brane contributions shift the positions of the vacua so that one has to analyze the full moduli space in this model to check whether dS vacua in a trustworthy regime could exist. There are of course also many more models that one can study in this new context. We leave a more detailed analysis to the future \cite{toappear}.

\subsection{The isotropic $S^3\times S^3/\mathbb{Z}_2\times \mathbb{Z}_2$ example}
\begin{figure}
    \centering
    \begin{subfigure}[b]{0.4\textwidth}
        \includegraphics[width=\textwidth]{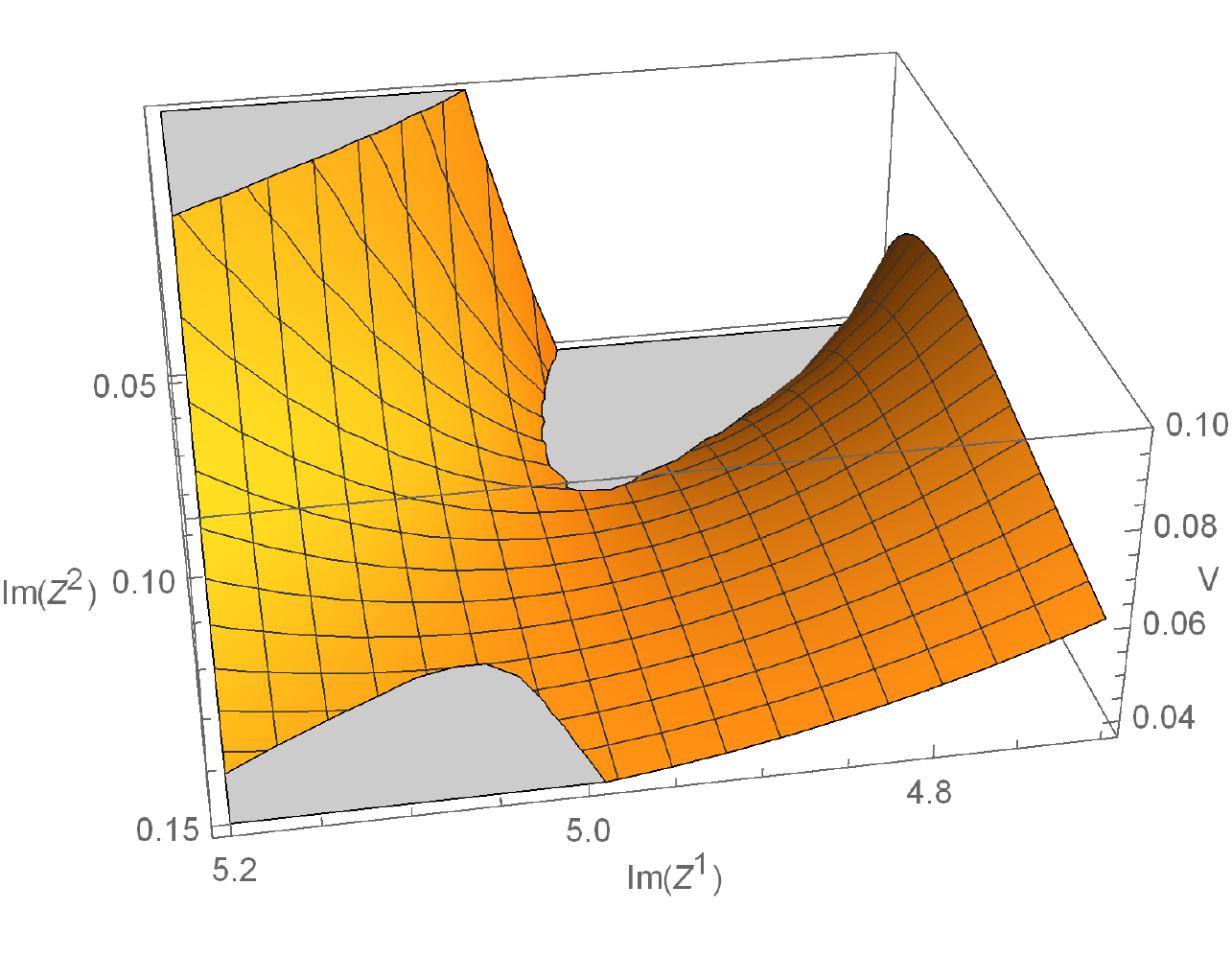}
    \end{subfigure}
    $\qquad\qquad$
    \begin{subfigure}[b]{0.45\textwidth}
        \includegraphics[width=\textwidth]{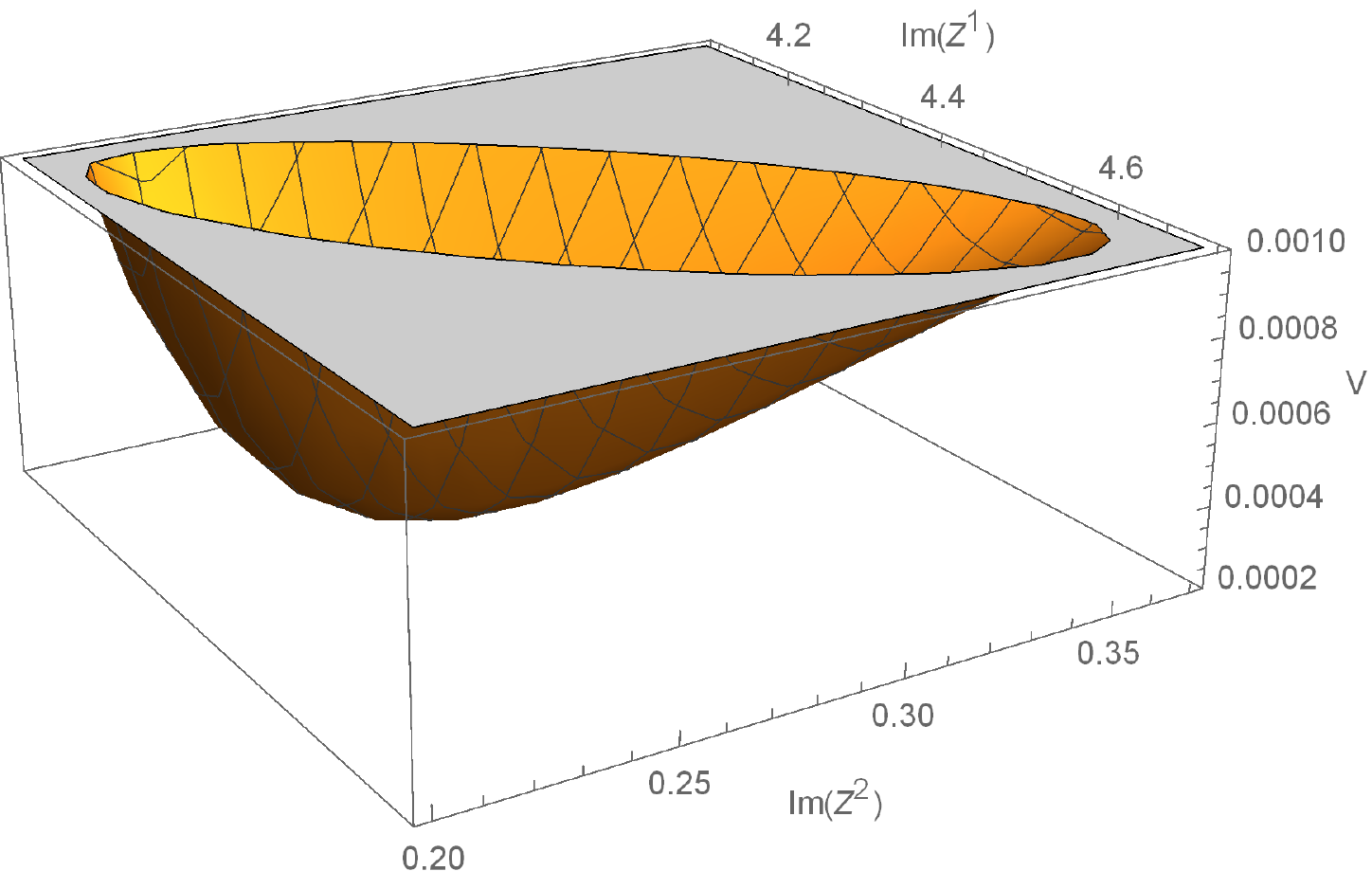}
    \end{subfigure}
    \caption{On the left we see that there is a tachyon in the plain spanned by (Im($Z^1$), Im($Z^2 $)). On the right the tachyon is gone, after we include the new term coming from anti-$D6$-branes.}
  \label{fig:notachyons}
\end{figure}

Probably the simplest example of compactifications of type II string theory is the compactification on $T^6/\mathbb{Z}_2\times \mathbb{Z}_2$, where one identifies the three $T^2$ in $T^6$. After this identification this model has only three complex moduli, whose imaginary parts correspond to a single volume modulus, a single complex structure modulus and the dilaton.\footnote{This model is the \emph{harmonic oscillator} of compactifications and is often called `STU-model' in the literature. We reserve $S$ for the nilpotent field here and also label the other moduli differently.} We compactify on this space and include in the NSNS sector H-flux, denoted by $h$ below, as well as metric fluxes. The latter are being equivalent to adding curvature and we choose them in such a way that the internal space is actually $S^3\times S^3$. This model has been studied in \cite{Villadoro:2005cu, Camara:2005dc, Aldazabal:2007sn, Flauger:2008ad, Caviezel:2008tf, Danielsson:2009ff, Danielsson:2010bc, Danielsson:2012et, Blaback:2013fca, Junghans:2016uvg, Junghans:2016abx, Roupec:2018fsp}. The only non-trivial fluxes we can add in the RR sector are $F_0$ and $F_2$ fluxes, whose parameter we denote by $f_0$ and $f_2$. Furthermore, we do an O6-orientifold projection and now allow for the addition of $N_{\overline{D6},K}$, $K=1,2$ anti-$D6$-branes on the two even 3-cycles. In our notation the \K and superpotential take the form 
\ba
K &=&-\ln\ls-2 \rmi (Z^1-\bar Z^1)\rs-3\ln\ls-2 \rmi(Z^2-\bar Z^2)\rs -\ln\ls\rmi(t-\bar t)^3-2 \rmi \frac{S\bar S}{N_{\overline{D6},K}(Z^K-\bar{Z}^K)}\rs \,,\cr
W &=&(h+3t) Z^1 -3(h-t)Z^2+ 3 f_2 t^2 - f_0 t^3+\mu^2 S\,.
\ea
In this model there are no D-terms. The internal volume is $8 \V_6 =\rmi(t-\bar t)^3$ and the four dimensional dilaton is $e^{-4{\phi_4}} = \e^{-4\phi} \V_6^2=2^8\text{Im}(Z^1)\text{Im}(Z^2)^3$. We have used $S^2=0$ to rewrite 
\be
-3\ln\ls-\rmi(t-\bar t)\rs +2\frac{S\bar S}{(t-\bar t)^3 N_{\overline{D6},K}(Z^K-\bar{Z}^K)}= -\ln\ls\rmi(t-\bar t)^3-2\rmi \frac{S\bar S}{N_{\overline{D6},K}(Z^K-\bar{Z}^K)}\rs.
\ee
The scalar potential of this model is not too complicated and we have actually been able to minimize it analytically in terms of the parameters. We have found that for suitable chosen values of the parameters we do indeed find stable dS solutions in our truncated model, i.e. the addition of anti-$D6$-branes has removed the tachyon, see figure \ref{fig:notachyons}.

We have explicitly checked that no other of the left-invariant moduli directions are tachyonic and that there are indeed metastable dS solutions in this truncation. There is a large parameters space and we leave it to the future \cite{toappear} to map it out and check whether one finds stable dS vacua in a trustworthy regime.\footnote{In addition to the quantization conditions one might worry that one could need on some cycles D6-branes as well as anti-D6-branes in order to find  dS vacua. This would probably lead to instabilities once one includes open string moduli. However, in this example one can find dS vacua for $N_1\neq0$, $N_2=0$, as well as $N_1=0$, $N_2\neq0$, so that this issue can be avoided.} One concrete set of values that leads to a dS a vacuum, that is however at small volume, strong coupling and does not have properly quantized fluxes and numbers of anti-$D6$-branes, is given by

\begin{center}
\begin{tabular}{ ccccc}
  $f_0=-1$ && $f_2=1$ && $h\approx -4.55$\\
  Re($t) \approx 1.14$ && Re$(Z^1) \approx -3.08$ && Re$(Z^2) \approx -0.519$ \\
  Im($t) \approx 1.76$ && Im$(Z^1) \approx 4.31$ && Im$(Z^2) \approx 0.251$ \\
  $N_1 =.05$ && $N_2 =.05$ && $\mu=1$\\
\end{tabular}
\end{center}

In this case the value of the scalar potential at the minimum is $V\approx 2\times 10^{-4}$ and the eigenvalues of the Hessian $\partial_i \partial_j V$, for $i,j$ running over real and imaginary part of $t$, $Z^1$ and $Z^2$ are approximately $\{ {4.7, 3.1, 0.95, 0.74, 0.024, 0.00012}\}$. This example is clearly not yet a full-fledged string theory solution. However, it is a proof of principle that the usefulness of uplift terms from anti-$Dp$-branes and the corresponding dS supergravity theories are very interesting and extend well beyond the KKLT \cite{Kachru:2003aw} and LVS \cite{Balasubramanian:2005zx} scenarios.

\section{Discussion}
In supersymmetry preserving compactifications of string theory, without the so-called pseudo-calibrated anti-$Dp$-branes, it seems  difficult to find de Sitter vacua. Here we have shown that the familiar uplifting term from anti-$D3$-branes can be generalized to all pseudo-calibrated anti-$Dp$-branes, wrapped on supersymmetric cycles. In particular, we have shown that including anti-$Dp$-branes in compactifications of type II string theory on SU(3)-structure manifolds leads to new positive terms in the scalar potential. We have explicitly calculated these terms and derived how they fit into the context of dS supergravity, which is the standard 4d $\N=1$ supergravity coupled to a nilpotent chiral multiplet (and potentially other constrained multiplets).

In type IIB it is known that to stabilize moduli in AdS space, in the first place, one should add some non-perturbative effects like D-brane instantons or gaugino condensation, which add to the potential some exponential dependence on moduli, instead of polynomial. Only when the moduli are stabilized in AdS using non-perturbative effects, an uplift via an anti-$D3$-brane/a nilpotent multiplet becomes useful and dS vacua are available. From this experience with anti-$D3$-branes we might expect that one may also needs non-perturbative effects for anti-$D5$, anti-$D7$, anti-$D9$-branes, to find some metastable dS vacua.\footnote{Similarly to the anti-$D3$-brane case, one probably also needs to substantially warp down the anti-brane tension, which is certainly problematic for the anti-$D9$-branes.} This needs to be studied but it seems possible that instantons/gaugino condensate can lead to a strong stabilization of all moduli, similar to the setup studied in \cite{Kallosh:2004yh}. General supergravity models of such type where studied in section 5.1 of \cite{Kallosh:2014oja}, where the Polonyi field $C$ would now be replaced by the nilpotent multiplet $S$ that arises from anti-branes in string theory. Based on these results, it seems possible that the new uplifting terms can supply new classes of dS vacua.

In type IIA flux compactifications moduli can be stabilized without using perturbative or non-perturbative corrections, but so far all dS vacua appearing in the standard 4d $\N=1$ supergravity had one universal tachyon. For dS vacua that are close to a no-scale Minkowski vacuum, the existence of such a tachyon in these models was proven in \cite{Junghans:2016abx}. Here we were up to a surprise in the simple STU-model: 
when the effect of the anti-$D6$-brane is taken into account via a nilpotent multiplet, the uplift removes the universal tachyon and all moduli in the truncation to left-invariant forms are stabilized in a dS vacuum. It remains to be seen, if full-fledged string theory solutions at large volume and small string couplings are available in this new setting. 

\

In conclusion, arguably, the discovery of dark energy may be viewed as a discovery of the Volkov-Akulov non-linearly realized supersymmetry from the sky.

\

\vspace*{.4cm}
\noindent
{\bf Acknowledgments:} 
We would like to thank D. Freedman, D.~Junghans, S. Kachru,  A.~Linde, M. Scalisi and T.~Van~Riet for useful discussions. We are grateful to Johan Bl{\aa}b\"ack for pointing out typos in the numerical solution in section 4.1 of earlier versions of this paper. RK is supported by SITP,  by the NSF Grant PHY-1720397, and by a Simons Foundation grant. TW is supported by an FWF grant with the number P 30265 and he is grateful for the hospitality of the Stanford Institute of Theoretical Physics where this work was performed.

\appendix
\section{Explicit four dimensional supergravity theories}\label{app:explicit}
In this appendix we discuss the detailed form of type II compactifications on SU(3)-structure manifolds, where we will use the notation of \cite{Robbins:2007yv}, to which we refer the interested reader for more details.  For simplicity, we restrict ourselves here to the four dimensional data without including open string moduli from D-branes (see for example \cite{Martucci:2006ij} and references to that article for how open string moduli appear). Our expressions are still correct in the presence of D-branes, if one sets the world volume fields like for example the scalar fields that control the D-brane position and the world volume vectors to zero.

Note, that we restrict to purely geometric compactifications so that the corresponding type II supergravity setups and their solutions should correspond to full-fledged string theory solutions, if they are found at large volume and weak coupling and satisfy the proper flux quantization and tadpole conditions.

\subsection{Type IIA with $O6$-planes}
In type IIA we have the RR-fluxes $F_0$, $F_2$, $F_4$ and $F_6$ that can thread the six dimensions of the SU(3)-structure manifold. In this sector we get scalar fields from the RR-form $C_3$ only since SU(3)-structure manifolds have no 1- and 5-cycles. However, they do have 2-cycles so that we can get also Abelian vector fields from $C_3$. The reduction of type IIA on CY$_3$ manifolds was worked out in \cite{Grimm:2004ua}. There it was found that the complex four dimensional scalar fields $Z^K$ and $t^a$ are given via the expansion of 
\ba
\Omega_c &\equiv& C_3 +2 \rmi e^{-\phi_4} \text{Re} (\Omega) = 2 Z^K a_K\,,\cr
J_c &\equiv& B+\rmi J = t^a \omega_a\,,
\ea
where $e^{-\phi_4}$ is related to the dilaton $e^\phi$ and the overall volume $\V_6=\frac{1}{3!} \int J \w J\w J$ via $e^{-\phi_4}= e^{-\phi}\sqrt{\V_6}$.

The resulting K\"ahler and superpotential was worked out for CY$_3$ manifolds in \cite{Grimm:2004ua}, while \cite{Villadoro:2005cu} was the first paper to describe an extra term in $W$ that arises from metric fluxes and \cite{Ihl:2007ah} showed that also non-vanishing D-terms can arise in this case. The four dimensional data is given by
\ba
K &=& 4\phi_4 -\ln(8 \V_6)\,,\cr
W &=&\int \lp \Omega_c \w (H -\omega\cdot J_c )+ F_6+F_4 \w J_c + \frac12 F_2 \w J_c\w J_c + \frac16 F_0 J_c \w J_c\w J_c\rp\,,\cr
f_{\a\b} &=& \rmi \hat{\k}_{a\,\a\b}t^a\,,
\ea
where we have given a simple form of $K$ that is an implicit function of the complex scalars. The curvature determines $\omega$ that maps the 2-form $J_c$ to a 3-form $\omega\cdot J_c$. The triple intersection numbers $\hat{\k}_{a\,\a\b}$ are determined by the geometry. The D-terms are given by
\be
D_\a = 2 \rmi e^{\phi_4} \hat{r}_\a^K {\cal F}_K\,,
\ee
where $\hat{r}_\a^K$ is determined by the curvature of the SU(3)-structure manifold and the ${\cal F}_K$ are purely imaginary functions of Re$(\Omega)$. Thus we find that type IIA reduced on SU(3)-structure manifolds in the presence of $O6$-planes gives generically rise to a standard four dimensional $\N=1$ supergravity theory that has F-terms as well as D-terms. 

The interesting feature of these type IIA flux compactifications is that all moduli can be stabilized by using the \emph{classical} scalar potential (as was first observed in \cite{Villadoro:2005cu, DeWolfe:2005uu, Camara:2005dc}). This means that perturbative and non-perturbative corrections can be neglected in solutions that are found at large volume and small coupling. This makes this class of models particular simple. It was proven in \cite{Hertzberg:2007wc} that in the absence of curvature, i.e. if the internal space is a Calabi-Yau 3-fold, then the scalar potential has only AdS vacua. If one includes curvature, then one can find dS critical points \cite{Flauger:2008ad, Caviezel:2008tf, Danielsson:2009ff}, however, until now all of these dS solutions had always one tachyonic direction, i.e. they were saddle points rather than local minima.

\subsection{Type IIB with $O3/O7$-planes} 
The RR-sector for type IIB compactified on SU(3)-structure manifolds gives only rise to parameters in the scalar potential via the $F_3$-flux (since there are no 1- and 5-cycles that we could thread with fluxes). The complex scalars that appear holomorphically in the superpotential are the complex expansion coefficients of the holomorphic 3-form $\Omega$ as well as $\tau$, $G^a$ and $T_\alpha$, defined via
\ba
\tau &=&C_0 + \rmi e^{-\phi}\,,\cr
G^a \omega_a &=& C_2 + \tau B\,,\cr
T_\a \mu^\a &=& C_4+ C_2 \w B +\frac12 \tau B\w B -\frac{\rmi}{2} e^{-\phi} J \w J\,.
\ea
The resulting four dimensional K\"ahler and superpotential have been worked out in \cite{Grimm:2004uq, Benmachiche:2006df} and are given by
\ba
K &=& - \ln\ls \rmi \int \Omega \w \bar\Omega\rs -4 \ln\ls-\rmi(\tau-\bar\tau)\rs - 2\ln\ls2\V_6\rs \,,\cr
W&=& \int \ls F_3+\tau H_3 +\omega \cdot (C_2+\tau B)\rs \w \Omega\,.
\ea
Here we have again given $K$ as implicit function of the moduli and $\omega \cdot (C_2+\tau B)$ is a 3-form that depends on the $G^a$ moduli as well as the curvature of the SU(3)-structure manifold. The volume is defined as in type IIA above via $\V_6=\frac{1}{3!} \int J \w J\w J = \frac{1}{3!}\k_{\a\b\gamma}v^\a v^\b v^\gamma$. The gauge kinetic function is a holomorphic function that depends only on the complex structure moduli contained in $\Omega$ and is given in equation (3.11) of \cite{Robbins:2007yv}. The D-terms are only non-vanishing, if there is curvature, encoded in $\hat{r}_{\a K}$, and they are given by
\be
D_K = -\frac{e^\phi}{2\V_6} \hat{r}_{\a K} v^\a\,.
\ee
So contrary to the F-term potential that satisfies a no-scale condition and depends on the K\"ahler moduli $T_\a$ only via an overall factor, the D-term potential can have a more interesting dependence on the volume via the dependence on $v^\a$. 

Generically, but also specifically in the case of Calabi-Yau compactifications, which have no curvature, one finds that the above classical scalar potential is not sufficient to stabilize all moduli. The flat directions are then lifted by perturbative and non-perturbative contributions. The most well studied scenarios here are KKLT \cite{Kachru:2003aw}, where a non-perturbative contribution $W_{np}$ is used and the LVS scenario \cite{Balasubramanian:2005zx}, where $W_{np}$ as well as a perturbative contribution to $K$ are used. All of these perturbative and non-perturbative corrections that modify the functions in the standard 4d $\N=1$ supergravity should be included and do not affect our general observation below that anti-D-branes give rise to a simple extra term in $K$ and $W$. 

\subsection{Type IIB with $O5/O9$-planes} 
In the case of an O5/O9 orientifold projection the story is similar to the case above. We again only have $F_3$ flux giving rise to parameters in the superpotential. However, the particular combination of fields that appear holomorphically in the superpotential are different. They are again given by the complex expansion coefficients of $\Omega$ and by $t^\a$, $L_a$ and $\T$ defined via
\ba\label{eq:moduliO5O9}
t^\a \mu_\a &=& C_2+\rmi e^{-\phi} J\,,\cr
L_a \tilde{\omega}^a &=& C_4+C_2\w B+\rmi e^{-\phi} B\w J\,,\cr
\T  &=& \int_{CY_3} \ls C_6+C_4\w B+\frac12 C_2\w B\w B+\rmi e^{-\phi} \lp \frac12 J\w B\w B -  \frac16 J \w J \w J\rp\rs\,.
\ea
The resulting four dimensional \K and superpotential have been worked out \cite{Grimm:2004uq, Benmachiche:2006df} and are given by
\ba
K &=& - \ln\ls \rmi \int \Omega \w \bar\Omega\rs - 4\ln\ls\e^{-\phi} \rs - 2\ln\ls8\V_6\rs \,,\cr
W&=& \int \ls F_3+\omega \cdot (C_2+\rmi e^{-\phi} J)\rs \w \Omega\,.
\ea
The K\"ahler potential is actually the same as for the O3/O7 orientifold projection but it is now an implicit function of the moduli given in equation \eqref{eq:moduliO5O9}. The gauge kinetic function depends holomorphically on the complex structure moduli contained in $\Omega$ and is given in equation (3.36) of \cite{Robbins:2007yv}. The H-flux does not appear in $W$ but appears in the D-terms via its expansion coefficients $p_k$. The D-terms are given by
\be
D_k =\frac{e^\phi}{2\V_6}\lp r_{ak} u^a -p_k\rp\,,
\ee
where the $u^a$ are the expansion coefficients of $B=u^a \mu_a$ and $r_{ak}$ is again determined by the curvature of the SU(3)-structure manifold.

Again one expects in this setup, like in any 4d $\N=1$ supergravity theory, that all flat directions will receive important perturbative and/or non-perturbative corrections. It would be very interesting to investigate whether the new terms we derive in this paper are sufficient (maybe together with quantum corrections) to lead to dS vacua in this setup.

\bibliographystyle{JHEP}
\bibliography{refs}

\end{document}